\newif\ifpreprint
\preprintfalse
\preprinttrue

\ifpreprint
	\documentclass[preprint2]{aastex}
\else
	\documentclass{aastex}
\fi

\usepackage{epsf}

\slugcomment{accepted by ApJ Letters}
\shorttitle{}
\shortauthors{Walter Dehnen}

\newcommand{\ssy}       {\scriptstyle}
\newcommand{\dsy}       {\displaystyle}
\newcommand{\tsy}       {\textstyle}
\newcommand{\B}[1]      {\mbox{\boldmath$#1$}}
\newcommand{\Bs}[1]     {\mbox{\boldmath$\ssy#1$}}
\newcommand{\bX}        {\B{X}}
\newcommand{\bY}        {\B{Y}}
\newcommand{\bZ}        {\B{Z}}
\newcommand{\bx}        {\B{x}}
\newcommand{\by}        {\B{y}}
\newcommand{\bR}        {\B{R}}
\newcommand{\beq}       {\begin{equation}}
\newcommand{\eeq}       {\end{equation}}
\newcommand{\ben}       {\begin{eqnarray}}
\newcommand{\een}       {\end{eqnarray}}
\newcommand{\bea}       {\begin{array}}
\newcommand{\eea}       {\end{array}}
\newcommand{\eqb}[1]    {(\ref{#1})}
\newcommand{\eqn}[1]    {equation \eqb{#1}}
\newcommand{\fig}[1]    {Fig.~\ref{fig:#1}}
\newcommand{\Fig}[1]    {Figure~\ref{fig:#1}}
\newcommand{\cD}[1]     {{\cal D}^{(#1)}}

\newcommand{\rA}        {{\rm A}}
\newcommand{\rB}        {{\rm B}}

\newcommand{\half}      {{\tsy{1\over2}}}
\newcommand{\sixth}     {{\tsy{1\over6}}}

\begin{document}

\title  {A Very Fast and Momentum-Conserving Tree Code}
\author {Walter Dehnen}
\affil  {Max-Planck-Institut f\"ur Astronomie, 
         K\"onigstuhl 17, D-69117 Heidelberg, Germany}
\email  {dehnen@mpia-hd.mpg.de}

\begin{abstract} \ifpreprint\noindent\fi
  The tree code for the approximate evaluation of gravitational forces is
  extended and substantially accelerated by including mutual cell-cell
  interactions. These are computed by a Taylor series in Cartesian coordinates
  and in a completely symmetric fashion, such that Newton's third law is
  satisfied by construction and hence momentum exactly conserved. The
  computational effort is further reduced by exploiting the mutual symmetry of
  the interactions. For typical astrophysical problems with $N$=$10^5$ and at
  the same level of accuracy, the new code is about four times faster than the
  tree code. For large $N$, the computational costs are found to scale almost
  linearly with $N$, which can also be supported by a theoretical argument, and
  the advantage over the tree code increases with ever larger $N$.
\end{abstract}


\section{Introduction} \ifpreprint\noindent\fi
The tree code (cf.\ Barnes \& Hut 1986, hereafter B\&H) has become an invaluable
tool for the approximate but fast computation of the forces in studies of
collisionless gravitational dynamics. It has been applied to a large variety of
astrophysical problems. The gravitational potential generated by $N$ bodies of
masses $\mu_n$ and at positions $\bX_{\!n}$ is
\beq
        \Phi(\bX) = - \sum_{n=1}^N\, \mu_n\; g(|\bX-\bX_{\!n}|),
\eeq
where $g(r)$ denotes the greens function, i.e.\ for un-softened gravity
$g(r)=G/r$. The essence of the tree code is to approximate this sum over $N$
terms by replacing any partial sum over all bodies within a single cell which is
well-separated from \bX\ by just one term. The inner structure of the cell is
partly taken into account using its multipole moments. This method reduces the
overall costs for the computation of all forces from ${\cal O}(N^2)$ to ${\cal
O}(N\log N)$.

The tree code, however, does not exploit the fact that the force due to the
contents of some cell is very similar at nearby positions (even though one may
use the fact that nearby bodies tend to have very similar interaction lists,
cf.\ Barnes 1990). Exploiting this is the idea of the fast multipole method
(FMM) \citep{FMM-a}. The FMM employs a (usually) non-adaptive structure of
hierarchical grids and considers only interactions between nodes on the same
grid level according to their geometrical neighbourhood. The gravitational field
due to some source cell and within some sink cell is approximated by a multipole
expansion in spherical harmonics, the order of which is adapted to meet
predefined accuracy limits. This method has been claimed to reduce the overall
amount of operations to ${\cal O}(N)$, but the tables given by
\citet{FMM-b} do not support this claim. \citet{CDM} find that the FMM needs
${\cal O}(N\log N)$ operations, and is significantly {\em slower\/} for
astrophysical applications than the tree code at comparable accuracy.

Instead of using a spherical multipole expansion of adaptive order, it is
actually more efficient to use a Cartesian expansion of fixed order. Moreover,
by preserving the symmetry of the gravitational interaction for mutual cell-cell
interactions, one can (i) reduce the computational effort and (ii) obtain a code
that satisfies Newton's third law by construction and hence results in exact
conservation of momentum, a property not shared by the traditional tree code.

\section{Description of the Code} \ifpreprint\noindent\fi
We start as the B\&H tree code with a hierarchical tree of cubic cells. Each
cell has up to eight sub-nodes corresponding to its octants. A node can be
either a single body or another cell. The tree-building phase (cf.\ B\&H) also
includes the computation of the cells' masses, centers $\bZ$ of mass, and
quadrupole moments.

\ifpreprint
\begin{figure}[t]
        \centerline{ \epsfxsize=70mm \epsfbox[51 162 585 461]{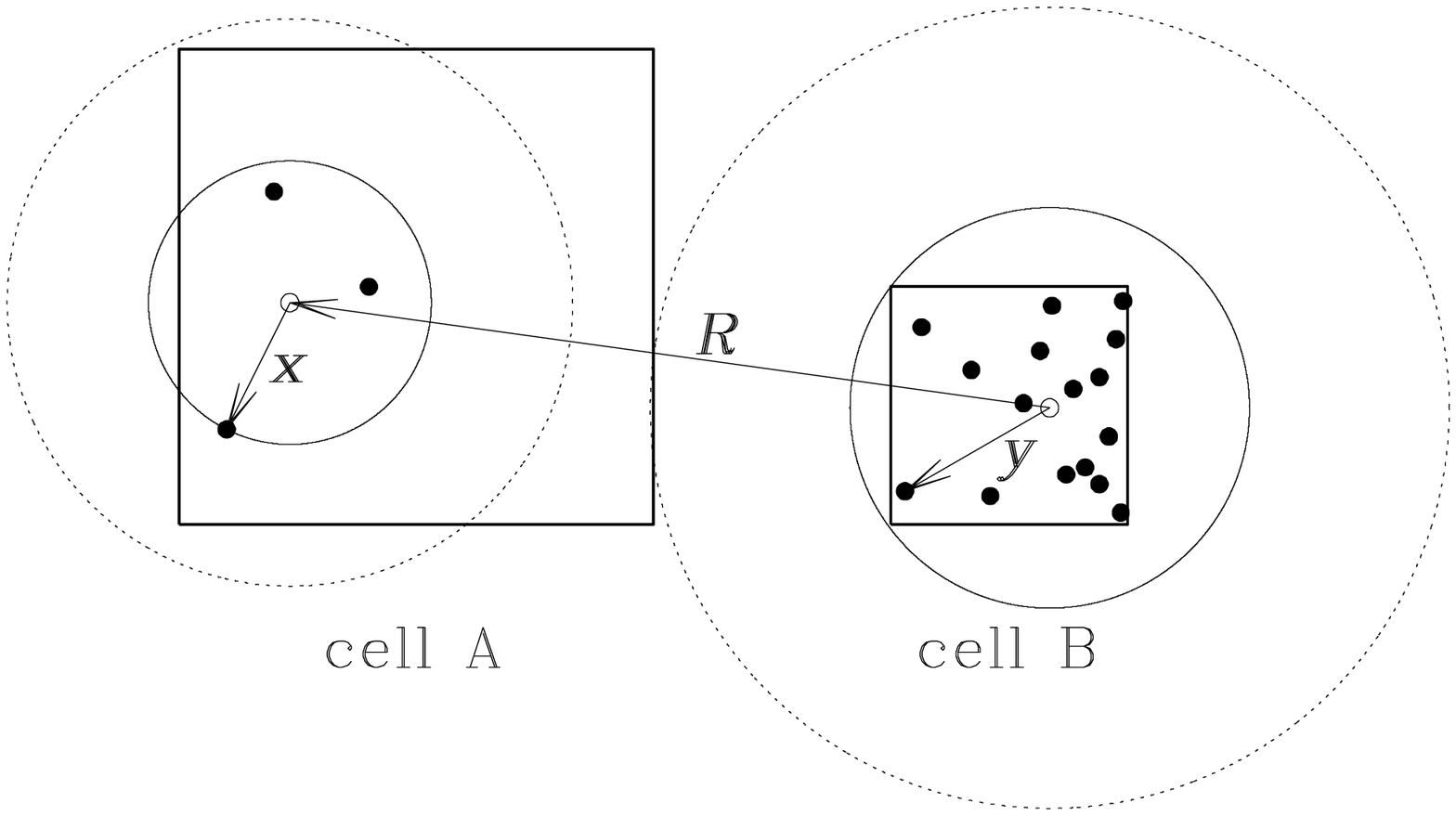}}
        \caption[]{\footnotesize Two well-separated cells. The solid and dotted
        circles have radii $r_{\rm max}$ and $r_{\rm max}/\theta$, respectively.
        \label{fig:example}}
\end{figure}
\fi
\subsection{The Opening Criterion} \ifpreprint\noindent\fi
In order to benefit from the symmetry of the gravitational interaction, the
opening criterion, which decides whether or not two nodes are well-separated so
that a direct mutual interaction is acceptable, must be symmetric, too. We
employ an extension of the criterion used in the tree-code: nodes A and B are
well-separated if
\beq \label{crit}
        |\bZ_\rA - \bZ_\rB| > (r_{\rm max\,A} + r_{\rm max\,B})/\theta,
\eeq
where the opening angle $\theta$ controls the accuracy of the code. $r_{\rm
max}$ is the radius of a sphere centered on the node's center of mass and
encircling all bodies within it. Bodies naturally have $r_{\rm max}\equiv0$,
i.e.\ two bodies are always well separated, while for the interaction between a
body and a cell the criterion \eqb{crit} reduces to that used in the tree code.
Note that if one additionally to \eqn{crit} requires $r_{\rm max\,A}=0$, the
standard tree code is re-covered, but the symmetry between A and B is broken.

\placefigure{fig:example}

There exist two upper limits for the radius $r_{\rm max}$. One is the distance
$b_{\rm max}$ between the cell's center of mass, \bZ, and its most distant corner
\citep{SW}. The other is
\beq
        \max_{{\rm sub-nodes}\; i}\left\{r_{{\rm max}\,i} + |\bZ_i-\bZ|\right\}
\eeq
\citep{BBCP}.  After computation of both these upper
limits, we take the smaller one to be $r_{\rm max}$. For cells with only a few
bodies like cell A in \fig{example}, the latter often gives values significantly
smaller than $b_{\rm max}$, while for cells with many bodies, like cell B in
\fig{example}, $b_{\rm max}$ is the tighter limit.

\ifpreprint
\begin{figure*}[t]
        \centerline{ \epsfxsize=150mm \epsfbox[68 539 580 706]{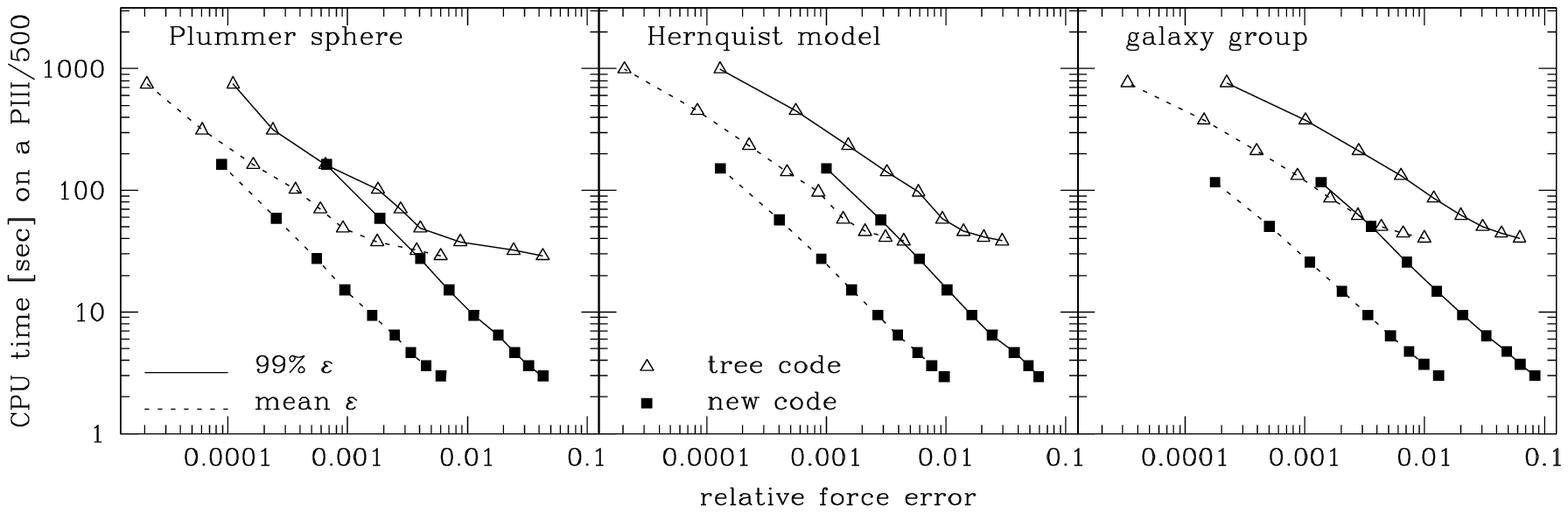}}
        \caption[]{\footnotesize CPU time consumption plotted versus the mean
        and 99\% relative force error for the three test cases. The dots
        correspond, from left to right, to
        $\theta=0.2,\,0.3,\,0.4,\,0.5,\,0.6,\, 0.7,\,0.8,\,0.9$ and 1.
        \label{fig:errors}}
\end{figure*}
\fi
\subsection{Approximating Gravity} \ifpreprint\noindent\fi
Consider two bodies at \bX\ and \bY\ which reside in two well-separated cells A
and B with centers of mass at, respectively, $\bZ_\rA$ and $\bZ_\rB$ and
separation $\bR\equiv\bZ_\rA-\bZ_\rB$. We may re-write $\bX-\bY= \bR+ (\bx-\by)$
with $\bx\equiv\bX-\bZ_\rA$ and $\by\equiv\bY-\bZ_\rB$ being small in magnitude
compared to $\bR$ (because the cells are well-separated, cf.\ \fig{example}).
The Taylor expansion of $g(|\bX-\bY|)$ around $\bR$ reads
\beq \label{Taylor}
        g(|\bX-\bY|) = \sum_p {1\over p!}\,\Big[\left[(\bx-\by)\cdot\B\nabla
        \right]^p g(|\B{r}|)\Big]_{\Bs{r}=\Bs{R}}.
\eeq
Separating powers of $x$ from powers of $y$ in \eqn{Taylor} and subsequently
taking the mass weighted sum over cell B yields a Cartesian multipole expansion
of the potential $\Phi_{\rB\to\rA}(\bX)$ at any position \bX\ within cell A and
due to all bodies inside cell B \citep{WS}. Since $\bZ_\rB$ was chosen to be the
center of mass of cell B, its dipole vanishes. The highest-order multipole
occurring in such an expansion may actually be omitted, since it only
contributes a constant to the approximation for $g$ and does not affect the
approximation for $\B\nabla g$ and hence the force. The expression of third
order thus reads (without the octopole; using Einstein's sum convention)
\beq \label{local-expn} \arraycolsep0.1em \bea{rlcl}
        \Phi_{\rB\to\rA}(\bX) &\approx M_\rB\bigg\{ 
      & &\cD0 + \half \tilde{Q}_{\rB\,ij} \cD2_{ij} 			\\
     &&+& x_i \Big[\cD1_i + \half\tilde{Q}_{\rB\,jk}\cD3_{ijk}\Big] 	\\
     &&+& \half x_ix_j\cD2_{ij} + \sixth x_ix_jx_k\cD3_{ijk}\bigg\},
\eea \eeq
where $M_\rB$ and $\tilde{Q}_\rB$ are the mass and {\em specific}
quadrupole moment
\beq
        \tilde{Q}_{\rB ij} \equiv {1\over M_\rB} 
        \sum_{\Bs{y}_n+\Bs{Z}_\rB\in\rm cell\;B} \mu_n\;y_{n\,i}\,y_{n\,j}
\eeq
of cell B, while $\cD{n}\equiv\B\nabla^n g(r)|_{r=|\Bs{R}|}$, i.e.\
\beq    \bea{r@{\;}c@{\;}l}
        \dsy    \cD0      &=& \dsy D^0,                                 \\      
                \cD1_i    &=&  R_i\,D^1,                                \\
                \cD2_{ij} &=& \delta_{ij}\, D^1(R) + R_iR_j\,D^2,       \\
        \dsy    \cD3_{ijk}&=& \dsy\left(\delta_{ij}R_k + \delta_{jk}R_i +
                                  \delta_{ki}R_j\right)D^2 + R_iR_jR_k\,D^3
\eea\mbox{\hspace{-8mm}}\eeq
with
\beq
        D^n\equiv \left.
        \left({1\over r}{\partial\over\partial r}\right)^n
        g(r) \right|_{r=|\Bs{R}|}.
\eeq
The symmetry between \bx\ and \by\ at every order of the Taylor expansion in
\eqn{Taylor} has two important consequences. First, if this expansion is
used to compute both $\Phi_{\rB\to\rA}(\bX)$ and $\Phi_{\rA\to\rB}(\bY)$,
Newton's third law is satisfied by construction.  Note, that our omission of the
octopole term broke the symmetry only in the zeroth order and has no effect on
the forces.

Second, the expressions for $M_\rA\Phi_{\rB\to\rA}$ and $M_\rB\Phi_{\rA\to\rB}$
are very similar: the expansion coefficients of second and third order differ
only by mere signs, such that computing these coefficients for both Taylor
series at one time is substantially faster than computing them at different
times.

The transformation, or shifting, of the expansion center to some other position
is trivial compared to the analogous procedure in the FMM (see Cheng et~al.).

\subsection{The Algorithm} \label{sec:algo} \ifpreprint\noindent\fi
The standard tree code computes the forces on each body by a recursive tree
walk, which visits each node exactly once as a gravity sink, and thus exhibits
an inherent asymmetry between sources and sinks. The new algorithm avoids this
asymmetry.

First, in the interaction phase, the Taylor series coefficients are evaluated
and accumulated in data fields associated with each node. This phase is based on
the concept of {\em mutual interactions\/} (MIs), pairs of nodes, A and B, such
that bodies in node A must receive forces from all bodies in node B, and vice
versa. We start by the MI describing the root-root self-interaction, and process
a given MI as follows.  (1) A body self-interaction is ignored; (2) a cell
self-interaction is split into the MIs between the sub-nodes\footnote{
	In a cubic oct-tree, these are at most 36 independent sub-MIs.},
and the process is continued on each of the new MIs; (3) a MI representing a
well-separated pair of nodes is executed: the Taylor coefficients are computed
and added to the nodes' corresponding data fields; (4) finally, in any other
case, the node with larger $r_{\rm max}$ is split, and up to eight new MIs are
created and processed.

Secondly, in the collection phase, the Taylor coefficients are passed down the
tree: the expansion center is shifted to the center of mass of the currently
active cell and the coefficients are accumulated. The Taylor expansion is
evaluated at the position of any body and the values for potential and
acceleration are added to its data fields (which may already contain
contributions accumulated during the interaction phase).

\section{Performance Tests} \ifpreprint\noindent\fi
We tested the new algorithm and compared it with the tree code in three typical
astrophysical situations: (1) a spherical Plummer model, representing a rather
homogeneous stellar system, (2) a spherical Hernquist-model (1990) galaxy, and
(3) a group of five such galaxies with various masses and scale radii. We
generated $10^5$ random initial positions from each of these cases, truncating
the density at 1000 scale radii, and evaluated the exact mutual forces at all
positions and the approximated forces due to the tree code (up to quadrupole
order) and the new code for opening angles $\theta$ between 0.2 and 1. We used
an optimally chosen softening with the biweight softening kernel (see Dehnen
2000), but the results are insensitive to these settings. Both approximate
methods have been coded by the author\footnote{
	At the same $\theta$, the tree code is twice as fast as a code publicly
	available from J.~Barnes, mainly because the new opening criterion leads
	to fewer interactions. However, even at the same number of interactions,
	the author's code was about 30\% faster.}
and use the same opening criterion (\S2.1). In order to measure the accuracy of
the approximated forces, we evaluated (cf.\ Capuzzo-Dolcetta \& Miocchi 1998)
\beq
	\varepsilon_n = |a_n - a_n^{PP}| \big/ a_n^{PP},
\eeq
where $a_n$ denotes the magnitude of the acceleration of the $n$th body due to
either of the approximate methods and $a_n^{PP}$ that of the exact computation.

\placefigure{fig:errors}

\Fig{errors} plots the CPU time needed for the force approximation on a 
Pentium III/500Mhz PC versus the mean relative error and that at the 99
percentile. For the new code, the time consumption scales almost inversely with
the error, while the tree code flattens off\footnote{
	This is, because of $r_{\rm max}$ is not proportional to the cell
	size, so that increasing $\theta$ from $0.7$ does not much decrease the
	number of interactions.}
at $\theta\ga0.7$. Evidently, at an acceptable level of accuracy, e.g.\
$\varepsilon_{99\%}=0.01$, the new code is about four times faster than the tree
code, even though it requires a smaller value of $\theta$ (0.5 as compared to
0.7 for the tree code).

\placefigure{fig:time}

For the test case of the Hernquist model, \Fig{time} plots the CPU time
consumption per body versus $N$ at fixed $\theta=0.7$ for the tree code and
$0.5$ for the new code. The tree code shows the well-known $\log N$
scaling, while for $N\ga10^5$ the new code requires only a constant amount of
CPU time per body. This can be explained as follows. Arranging eight root
cells to a new root box, increases $N$ to $8N$ and the number $N_I$ of
interactions to $8(N_I+N_+)$, where $8N_+$ interactions are needed to compute the
forces between the former root cells. Thus,
\beq \label{scal-diff}
	{{\rm d}N_I\over{\rm d}N} \simeq 
	{N_I\over N}{\Delta\log N_I\over\Delta\log N} = 
	{N_I + N_+/\ln8\over N}.
\eeq
In the tree code, $N_+\propto N$ yielding $N_I\propto N\log N$. In the new code,
$N_+$ may be estimated to contain two contributions, a constant term accounts
for the interactions with distant nodes, and a term $\propto N^{2/3}$ for those
on the {\em surface\/} of the former root cells. Inserting this into
\eqn{scal-diff} yields
\beq
	N_I \propto N - c_1 N^{2/3} - c_2
\eeq
with constants $c_1$ and $c_2$ that depend on $\theta$. Thus, at large $N$, a
linear relation is approached. Note that this argument differs from that given
for the FMM by \citet{FMM-a}, who assumed that the resolution may remain fixed
when increasing $N$.

\ifpreprint
\begin{figure}[t]
        \centerline{ \epsfxsize=70mm \epsfbox[50 303 580 706]{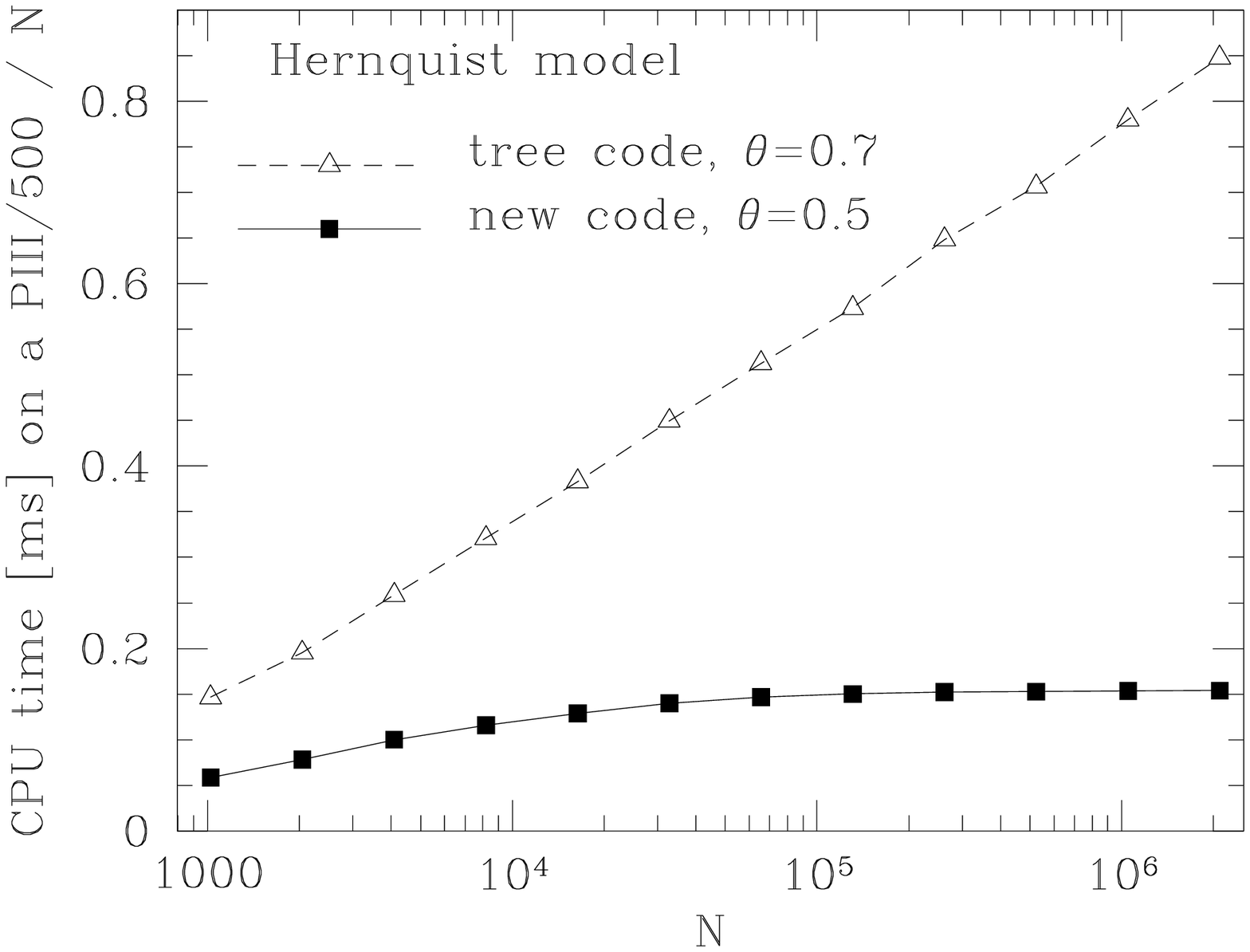}}
        \caption[]{\footnotesize CPU time per body plotted versus $N$ for test
        case 2 (Hernquist model) \label{fig:time}}
\end{figure}
\fi
\section{Discussion} \ifpreprint\noindent\fi
A new code for the approximate evaluation of gravitational forces has been
presented, tested, and compared to the tree code. This new code is substantially
faster than the tree code. Moreover, unlike the latter, it satisfies Newton's
third law by construction, such that any $N$-body code based on it will not
introduce spurious net-accelerations. The new code is based on a Taylor
expansion of the greens function in Cartesian coordinates and incorporates
mutual cell-cell interactions. The simple algorithm is well suited for
implementation on parallel computers: different mutual interactions (MIs) can be
passed to different CPUs.

The scaling of the CPU time required for the mutual forces of a number $N$ of
bodies becomes essentially linear at $N\ga10^5$, so that with ever larger $N$
the new code is increasingly faster than the tree code, allowing for a
substantial improvement in simulations employing large number of bodies. The
only disadvantage is the increased requirement of memory compared to the
standard tree code: 20 floating point numbers per cell are needed to hold the
Taylor expansion coefficients. (By using a tree-walking algorithm instead of
that given in \S\ref{sec:algo}, one can avoid this at the price of enhanced CPU
time consumption.)

In spirit, the new code is similar to Greengard \& Rokhlin's (1987) fast
multipole method, but is more efficient because it (i) uses a Cartesian instead
of a spherical harmonic multipole expansion and (ii) fixes the order of the
expansion while controlling the accuracy via the interaction condition, rather
than fixing the interactions and adapting the expansion order to the accuracy.

A concern with codes based on cell-cell interactions is their performance in the
presence of individual time steps. Clearly, when not all the forces are to be
computed, such codes fare less favorably. However, when the forces for all
bodies within some domain are desired, the new code is still a significant
improvement over the tree code.

The new code has been written in C++ and
\ifpreprint will be \else is \fi
electronically available from the author upon request.

\acknowledgments
The author thanks Tom Quinn for valuable discussions and Joshua Barnes, Lars
Hernquist, Junichiro Makino, Andy Nelson, and Thorsten Naab for useful comments.

\ifpreprint
  \def\thebibliography#1{\subsection*{References}
    \list{\null}{\leftmargin 1.2em\labelwidth0pt\labelsep0pt\itemindent -1.2em
    \itemsep0pt plus 0.1pt
    \parsep0pt plus 0.1pt
    \parskip0pt plus 0.1pt
    \usecounter{enumi}}
    \def\refpar{\relax}
    \def\newblock{\hskip .11em plus .33em minus .07em}
    \sloppy\clubpenalty4000\widowpenalty4000
    \sfcode`\.=1000\relax
    \footnotesize}
  \def\endthebibliography{\endlist}
\fi

\ifpreprint \relax \else
\clearpage \onecolumn

\begin{figure}\caption[]{
	Two well-separated cells. The solid and dotted
        circles have radii $r_{\rm max}$ and $r_{\rm max}/\theta$, respectively.
        \label{fig:example}}
\end{figure}

\begin{figure}\caption[]{
	CPU time consumption plotted versus the mean and 99\% relative force
	error for the three test cases. The dots correspond, from left to right,
	to $\theta=0.2,\,0.3,\,0.4,\,0.5,\,0.6,\, 0.7,\,0.8,\,0.9$ and 1.
	\label{fig:errors}}
\end{figure}

\begin{figure}\caption[]{
        CPU time per body plotted versus $N$ for test case 2 (Hernquist model)
        \label{fig:time}}
\end{figure}

\fi 
\end{document}

       	\clearpage
        \centerline{\epsfxsize=140mm \epsfbox[51 162 585 461]{Dehnen.fig1.ps}}
        \centerline{\fig{example}}

        \clearpage
        \centerline{ \epsfxsize=160mm \epsfbox[68 539 580 706]{Dehnen.fig2.ps}}
        \centerline{\fig{errors}}

        \clearpage
        \centerline{ \epsfxsize=120mm \epsfbox[50 303 580 706]{Dehnen.fig3.ps}}
        \centerline{\fig{time}}

\fi 

\end{document}